\documentclass[12pt]{article}

\usepackage{amsmath,amssymb,amsthm,amscd}
\usepackage{graphicx}
\usepackage{authblk}

\usepackage{natbib}
\usepackage{algorithm}	
\usepackage{algorithmic}
\usepackage{bm}
\usepackage{bbm}
\usepackage{enumitem}

\usepackage{listings}
\usepackage{setspace}
\usepackage{cprotect}

\usepackage[toc,page]{appendix}

\graphicspath{ {Images/} }

\theoremstyle{plain}

\usepackage{setspace}

\providecommand{\keywords}[1]{\textbf{\textit{Keywords: }} #1}

\begin{document}

\title{Scalable visualisation methods for modern\\ Generalized Additive Models}

\author[1,$\dag$]{Matteo Fasiolo}
\author[2]{Rapha\"el Nedellec} 
\author[2]{Yannig Goude}
\author[1]{Simon N. Wood}
\affil[1]{School of Mathematics, University of Bristol, United Kingdom.}
\affil[2]{\'Electricit\'e de France R\&D, Saclay, France.}
\affil[$\dag$]{Correspondence: matteo.fasiolo@bristol.ac.uk}

\maketitle

\begin{abstract}
In the last two decades the growth of computational resources has made it possible to handle Generalized Additive Models (GAMs) that formerly were too costly for serious applications. However, the growth in model complexity has not been matched by improved visualisations for model development and results presentation. Motivated by an industrial application in electricity load forecasting, we identify the areas where the lack of modern visualisation tools for GAMs is particularly severe, and we address the shortcomings of existing methods by proposing a set of visual tools that a) are fast enough for interactive use, b) exploit the additive structure of GAMs, c) scale to large data sets and d) can be used in conjunction with a wide range of response distributions. The new visual methods proposed here are implemented by the \verb|mgcViz| R package, available on the Comprehensive R Archive Network\footnote{The code for reproducing the results in the paper can be found at https://github.com/mfasiolo/code\_for\_GAM\_visual\_paper.}.
\end{abstract}

\vspace{4pt}

\keywords{Generalized Additive Models; visualisation; electricity load forecasting; residuals checking; regression modelling; interactive model building}

\section{Introduction}

The aim of this paper is to propose new visualisation tools for interactive model checking and development in smooth additive models, with a particular focus on large models for big data sets, and model checking beyond simple exponential family regression. In particular, recent computational developments in GAM fitting methods, such as \cite{wood2015generalized}, \cite{wand2017fast} and \cite{wood2017generalizedSmoke}, have made it possible to use these models to explore very large data sets. However, visual methods and software have lagged behind, to the point that, for data sets comprising over $10^6$ observations, GAM model fitting might take less time than rendering basic visual residuals checks. The methods described here address this issue by binning the data and summarising it into a form that can displayed effectively, as suggested by \cite{wickham2013bin}. To enable interactive exploration, the new tools aim at handling data sets comprising $10^7$ to $10^8$ observations within a few seconds.  

Beside faster computation, in the last two decades GAM methods have expanded in terms of the range of models that can be fitted. Indeed, modern GAMs have moved beyond the exponential family, and are not limited to modelling the mean. Here we consider additive models that can be fitted using the general framework of \cite{wood2016smoothing}. In particular, if we let $y$ be the response variable, then
$$
y_i \sim \mathcal{D}_m(y|{\bm \theta}_i), \;\;\;
g_k(\theta_{ki}) = \sum_{j \in \mathcal{S}_k} f_{kj}({\bf x}_i), \;\;\; \text{for} \;\; i = 1, \dots, n, 
$$
where $\mathcal{D}_m(y|{\bm \theta})$ is a distribution parametrized by ${\bm \theta}_i=\{\theta_{1i}, \dots, \theta_{pi}\}$, the $f_{kj}$'s are unknown smooth functions of the covariate vector $\bf x$, $g_1, \dots, g_p$ is a sequence of known smooth monotonic link functions and $\mathcal{S}_k$ is the set of indices specifying on which smooth effects $\theta_k$ depends. The $f_{kj}$'s are  constructed using basis expansions of low rank, such as splines, whose complexity is controlled using ridge penalties on the regression coefficients. This general framework includes Generalized Additive Models for Location, Scale and Shape (GAMLSS) by \cite{rigby2005generalized}, where all parameters of the response distribution can be modelled via additive functions of the covariates. 
While this permits greater flexibility, a practitioner is now left with the task of specifying several linear predictors, rather than one. Further, such models are often expensive to fit, which makes performing exhaustive or automated variable selection impracticable. Instead, this paper proposes a set of visual tools intended to aid visual variable selection. In particular, we propose several visualisations aimed at detecting residual patterns and anomalies which, importantly, quantify the uncertainty of the residual patterns and are applicable to most distributional GAMs.



We illustrate the new visualisations in the context of building a GAM model for predicting electricity demand on the UK grid. Here `trying all possible models' is infeasible and good visual tools are essential for interactive model building. Effective visualisation of such effects is also important in this context, as operational forecasters need to understand the estimated effects to assess their physical plausibility and judge when it may be safe to use the model despite unusual covariate configurations. Good visualisation is key to this, hence this application helps us motivating some new smooth effect plots, which permit visual uncertainty assessment and can be manipulated interactively.

We consider electricity demand data from www.nationalgrid.com, covering the period between January 2011 and June 2016, and containing 48 daily observations at 30min intervals. 
We integrate it with hourly temperatures from the National Centers for Environmental Information. 
We first consider a Gaussian GAM, where the expected load is modelled by
\begin{equation}\label{eq:gausModel}
\mathbb{E}(L_i) = \beta_0h_{d(i)} + \sum_{j=1}^7 \beta_j w^j_{d(i)} + \beta_8L_{i-48} + f_1(t_i) + f_2(T_i, I_i) + f_3(T^s_i, I_i) + f_4({\tt toy}_i, I_i),
\end{equation}
for $i=1,\dots,n$. Here $L_i$ is the $i$-th observed load, $L_{i-48}$ is its value in the same half-hourly period of the previous day, $d(i)$ is the date and $h_{d(i)}$ is equal to one if $d(i)$ is a bank holiday, zero otherwise. Similarly, $w^j_{d(i)}$ is equal to one if $d(i)$ is the $j$-th day of the week, and $\beta_0, \dots, \beta_8$ are unknown coefficients. $t_i$ is time since the 1st of January 2011 at 30min resolution, and $f_1$ is a smooth effect constructed using a cubic spline of rank six, meant to capture the long term trend. $T_i$ and $T_i^s$ indicate temperature and smoothed temperature, where the latter was obtained using $T^s_i=\alpha T_i+(1-\alpha)T^s_{i-1}$, with $\alpha = 0.05$, and it is meant to capture thermal inertia in buildings, that is the fact that it takes some time for external temperature to affect internal temperature. $I_i\in\{1,2,\dots,48\}$ is the half-hour interval of the day  and ${\tt toy}_i \in (0, 1)$ is the time of the year at half-hourly resolution. $f_2$, $f_3$ and $f_4$ are bivariate smooth functions, based on tensor products of rank 200, 200 and 600, built using cubic spline marginal bases for $T_i$ and $T^s_i$, and cyclic marginal bases for $I_i$ and ${\tt toy}_i$.

The rest of the paper is structured as follows. Section \ref{sec:software} discusses the software implementation of the methods proposed here, while sections \ref{sec:visModBuild} and \ref{sec:smoothEff} present, respectively, visual diagnostics and methods for smooth effect uncertainty visualization. Sections \ref{sec:loadEx1} and \ref{sec:loadEx2} focus on the load forecasting application, and show how the visualisations described in first part of sections \ref{sec:visModBuild} and \ref{sec:smoothEff} can help improving upon the Gaussian GAM described above.   

\section{Software} \label{sec:software}
The visual methods developed here could be applied to GAMs fitted with a variety of software, but for concreteness we focus on the \verb|mgcv| package in the \verb|R| statistical software. \verb|mgcv| provides tools for building and fitting GAM models including a wide variety of smooth or random effects and response distributions, and it is supplied by default with \verb|R|. In the last few years the flexibility and scalability of \verb|mgcv| has benefited from the inclusion of the fitting methods of \cite{wood2015generalized, wood2016smoothing, wood2017generalizedSmoke}. However, these improvements have not been matched by the development of adequate visual tools for checking and exploring the model output. The \verb|mgcViz| package is an extension of \verb|mgcv| meant to fill this gap, by offering scalable and interactive visual tools for model development and results presentation. 

Most of the visual tools in \verb|mgcv| are implemented by two functions: \verb|plot.gam|, which plots the smooth or parametric terms, and \verb|gam.check|, which performs model checking. Rather than providing few multiple purpose functions, \verb|mgcViz| exploits the additive structure of GAMs to set up a modular object-oriented framework, briefly outlined here. Let \verb|obj| be a fitted GAM model, that is the output of \verb|mgcv::gam|. Then \verb|obj<-getViz(obj)| will convert it to an object of class \verb|gamViz|, which can be visualised using \verb|mgcViz|. For example, the $k$-th fitted smooth effect contained in \verb|obj| can be extracted using \verb|fk<-sm(obj,k)|, and transformed into a visual object of class \verb|plotSmooth| by using \verb|plot(fk)|. The generic \verb|plot| function calls a specific plotting method, depending on the class of \verb|fk|. Parametric terms can be extracted using the \verb|pterm| function, and can then be plotted similarly. 

Most of the graphical objects produced by \verb|mgcViz| belong to the \verb|plotSmooth| class, and contain one or more objects of class \verb|ggplot|, defined in the \verb|ggplot2| package \citep{ggplot2}. This allows us to exploit the powerful layering system provided by \verb|ggplot2|, which enables users to superpose several graphical layers, possibly based on different data, on a single plot  \citep{wickham2010layered}. For example, if \verb|fk| is a standard one-dimensional smooth, then we can do
\begin{verbatim}
plot(fk) + l_dens(type = "cond") + l_fitLine() + l_ciLine(linetype = 2)
\end{verbatim}
which plots a heatmap representing the conditional density of the partial residuals overlaid by the fitted effect with 95\% confidence intervals (CIs) (for an example of the former, see Figure \ref{eq:gausModel}a). In \verb|mgcViz| all functions with prefix \verb|l_| output graphical layers, which can be added to the effect plots by using the overloaded \verb|+| operator. One advantage of this system, relative to using few multiple purpose plotting functions, is that it can be extended easily by adding new layering methods. Secondly, most graphical and algorithmic parameters are specified directly at the individual layer level, allowing more control than the plotting methods provided by \verb|mgcv|. Thirdly, basing the plotting system around \verb|ggplot| objects allows us to exploit the vast array of layers provided by \verb|ggplot2| and to automatically convert  the plots to \verb|plotly| objects \citep{sievert2017plotly}, which provide interactive features such as zooming and sub-setting, useful for exploring the model output and for diagnostic purposes.

In Section \ref{sec:visModBuild} we describe several new methods for GAM model checking and smooth effect visualisation. Each time we describe a visual tool, we detail its mathematical and algorithmic structure, and we provide a reference to its implementation in \verb|mgcViz|. This is because, while each new visualisation is useful individually, we argue that the layered object-oriented framework just outlined is essential for creating an extensible, user-friendly and easily maintainable visual toolbox for GAM modelling.

\section{Visual tools for interactive GAM model building} \label{sec:visModBuild}

\subsection{Scalable interactive QQ-plots for general GAMs} \label{sec:qqplots}

Our aim here is developing QQ-plot methods for GAMs that are sufficiently fast to permit interactive exploration even for large data sets, that provide non-asymptotic reference intervals around the QQ-curve and that generalise to almost any response distribution. 

Consider a data set consisting of covariate vectors ${\bf x}_1, \dots, {\bf x}_n$ and continuous responses $y_1, \dots, y_n$, whose conditional distribution has p.d.f. $p(y|{\bf x})$. For simplicity, assume that ${\bf x} \in \mathbb{R}^d$.  Let $p_m(y|{\bf x})$ be the model-based density, and define the residuals $r_i = t(y_i|{\bf x}_i)$, for $i=1,\dots,n$, where $t(y|{\bf x}_i)$ is a general sequence of transformations. Define the marginal density $p(r) = \int p(r|{\bf x})p({\bf x})d{\bf x}$ and its model-based estimate $\hat{p}_m(r) = n^{-1}\sum_i p_m(r|{\bf x}_i)$. Notice that, while $p_m(r|{\bf x})$ depends only on the model and on $t(y|{\bf x})$, $p(r|{\bf x})$ depends also on the data generating process. In a regression context, QQ-plots are typically used to compare the quantiles of $\hat{p}_m(r)$ with those of $p(r)$. The latter are typically unavailable, but can be estimated using the sample $r_i \sim p(r)$, for $i=1,\dots,n$. Under discrete $y$ the same definitions hold, but we would be dealing with probability mass functions, not densities.

The ease with which the objectives stated above can be achieved mainly depends on the transformation, $t(y_i|{\bf x}_i)$, and the tractability of $p_m(y|{\bf x})$. Common transformations are:
\begin{enumerate}[label=(\alph*)]
\item $F_m(y_i|{\bf x}_i)$, where $F_m(y|{\bf x})$ is the conditional c.d.f. corresponding to $p_m(y|{\bf x})$.
\item $\Phi^{-1}\{F_m(y_i|{\bf x}_i)\}$, where $\Phi$ is a standard normal c.d.f.. These are the `quantile' residuals of \cite{dunn1996randomized}.
\item $\{y_i - \mu_m({\bf x}_i)\} / \sqrt{v_m({\bf x}_i)}$, where $\mu_m({\bf x})$ and $v_m({\bf x})$ are model-based estimates of, respectively, $\mathbb{E}(y|{\bf x})$ and $\text{var}(y|{\bf x})$. This produces scaled Pearson residuals.
\item $\text{sign}\{y_i - \mu_m({\bf x}_i)\}\sqrt{d_i}$, where $d_i$ is the $i$-th deviance component. This choice leads to the deviance residuals.
\end{enumerate}  
Leaving aside the sampling variability of the estimated model coefficients, and under a continuous $y$ and a well specified model, options (a) and (b) should produce residuals that are, respectively, close to uniformly and normally distributed. 
In either case, the leading cost of computing the observed quantiles is $O(n\log n)$, if the $r_i$s are sorted sequentially. Under (a) reference intervals (RIs) can be obtained using the critical regions of the Kolmogorov-Smirnov statistic \citep{michael1983stabilized} while,    
under (b), $\alpha\%$ RIs around a normal quantile $z$, associated with probability p, can be approximated using $\pm\Phi^{-1}\{(1+\alpha)/2\}\phi(z)^{-1}\{p(1-p)/n\}^{1/2}$, where $\phi$ is a standard normal p.d.f. \citep{buuren2001worm}.

QQ-plots based on residuals (a) and (b) achieve the first two objectives stated above, but are difficult to interpret when $y$ takes few discrete values. Scaled Pearson and deviance residuals are generally continuous even when the response is discrete and are arguably more popular than uniform or quantile residuals. While, under a general response distribution, not much can be said about the distribution of Pearson residuals, \cite{pierce1986residuals} argue that deviance residuals are generally close to normally distributed, in an exponential family context. However, there are cases of practical importance, such as when $y$ consists of low counts, where the resulting QQ-plots shows deviations from a straight line, even when the model is correct \citep{ben2004quantile}.
Solutions to the discreteness issue are offered by \cite{dunn1996randomized}, who obtain continuous quantile residuals by randomising the uniform residuals $F_m(y_i|{\bf x}_i)$, and then transforming them to normality, while \cite{czado2009predictive} propose a non-randomized algorithm for producing uniform QQ-plots. While these methods might be included in future versions of \verb|mgcViz|, the current \verb|qq.gamViz| function addresses the issue by adopting the simulation-based approach of \cite{augustin2012quantile}. Briefly, it simulates $l$ $n$-vectors of responses from $p_m(y|{\bf x}_1),\dots,p_m(y|{\bf x}_n)$ using parameters fixed at their estimated value, transforms them to residuals, and compares the observed ordered residuals with their simulated counterparts. The advantage of this method is that it is very general (e.g. it can be applied to Pearson residuals) and that RIs for the model-based quantiles can be estimated using the simulations. The cost of computation becomes $O(ln\log n)$, but the $l$ iterations are independent and thus easy to parallelise.

Under residual types (a) and (b) \verb|qq.gamViz| requires few seconds to calculate the QQ-curve and its RIs for data sets of size $10^7$ on a single core. In a similar setting and with $l=100$, the simulation-based methods of \cite{augustin2012quantile} might take few minutes to produce the same output. The speed of the simulations could be improved, but at the time of writing this performance seems acceptable, if compared with the time needed to fit a GAM model to such a large data set. However, once the QQ-plot has been computed, it needs to be rendered graphically. Over-plotting is not an issue for QQ-plots, but \verb|R| plotting facilities slow down considerably for $n>10^6$, which impedes performing interactive actions on the plot. We address this problem by binning the points forming the QQ-plot and its RIs before rendering. In particular, we construct $b_0$ bins along the QQ-curve, with each bin covering the same arc-length. The arc-length of the original QQ-curve is $
h = \sum_{i=2}^n \{(r_i - r_{i-1})^2 + (\bar{r}_i - \bar{r}_{i-1})^2\}^{\frac{1}{2}},
$   
where the $r_i$'s and $\bar{r}_i$'s are the sorted observed and model-based residuals. Having defined $b_0$ bins along $h$, we assign each point to a unique bin, and average the $r_i$'s and $\bar{r}_i$'s belonging to each bin. The output are two sequences $s_j$ and $\bar{s}_j$ with $j = 1, \dots, b$, where $1 \leq b \leq b_0$ because some of the bins might be empty. 


The cost of binning is $O(n)$ or $O(ln)$ if one wants to plot all the simulated QQ-lines. Notice that, with $n=10^7$ and $l=10^2$, the latter option would require plotting $10^9$ points if binning is not used, which is infeasible even for non-interactive use. For data sets of this size binning takes less than a second, which permits interactive features such as zooming. In \verb|mgcViz| interactive zooming is provided by the \verb|shine.qqGam| method, which transforms the output of \verb|qq.gamViz| into a Shiny application \citep{shinyManual}. This interactive feature can be used to look at specific parts of the QQ-plot, without paying again the $O(n\log n)$ price implied by sorting. Adapting the bins to the new zooming area allows the user to check whether binning has hidden any feature of the original QQ-plot.

\subsection{Beyond QQ-plots: conditional residual checks} \label{sec:checks}

The QQ-plot methods described in Section \ref{sec:qqplots} focus on the marginal distribution of the residuals. Here we describe tools for assessing departures from the model-based conditional residuals distribution, along one or two covariates. Let $x_j$, with $j\in\{1,\dots,d\}$, indicate the $j$-th covariate and let $x_{ij}$, for $i=1,\dots,n$, be its observed values. Plotting the residuals against the $j$-th covariate allows visualisation of a sample from $p(r|x_j) \propto \int p(r,{\bf x})dx_1\cdots dx_{j-1}dx_{j+1}\cdots dx_{d}$ which, in a classical regression context, helps identifying outliers, important omitted variables, non-linearities, heteroscedasticity and autocorrelations \citep{cox1968general}. When working with general GAMs, such plots can also help assessing over or under-smoothing and can prompt the addition of effects that let the skewness or tail behaviour of the response distribution vary with the covariates. 

\verb|mgcViz| provides several methods for comparing the observed and model-based conditional distributions of the residuals, which have been implemented using the layered framework described in Section \ref{sec:software}. In particular, if \verb|obj| is an object of class \verb|gamViz| containing a fitted GAM model, then the function calls \verb|check1D(obj,"x1")| and \verb|check2D(obj,"x1","x2")| extract the residuals from fitted GAM model, and create graphical objects representing the relation between the residuals and the covariates $x_1$ and $x_2$. Visual residuals diagnostics can be then plotted by adding one of the layers described in the following.


The \verb|l_densCheck| layer produces a heatmap representing the  distance, $\delta_{p, p_m}(r|x_j)$, between $p(r|x_j)$ and $p_m(r|x_j)$. The user can provide any distance function, and Figure \ref{fig:diagGaus}c-d shows two examples where $\delta_{p, p_m}(r|x_j) = \{p(r|x_j)^{1/2} - p_m(r|x_j)^{1/2}\}^{1/3}$. $p(r|x_j)$ is estimated using $p(r|{x}_j) = p(r, x_j)/p(x_j)$, where fast kernel density estimates (k.d.e.) of $p(r, x_j)$ and $p(x_j)$ are computed using the \verb|KernSmooth| package \citep{wand2006kernsmooth}. This implements linear binning in one or two dimensions \citep{wand1994fast}, hence it scales well with $n$. In Figure \ref{fig:diagGaus}c-d $p_m(r|x_j)$ is analytically available but, when it is not, it is possible to simulate residuals from the model and to use them to estimate $p_m(r|x_j)$, as done for $p(r|x_j)$. The Supplementary Material contains few examples meant to help practitioners interpreting the output of \verb|l_densCheck|.  

Plotting $\delta_{p, p_m}(r|x_j)$ provides much detail regarding the residuals distribution, but practitioners are often interested in specific residuals patterns (e.g. heteroscedasticity). Such focused checks can be performed using the \verb|l_gridCheck1D| layer which assigns each residual to one of $b$ bins, equally spaced along $x_j$, and summarises the residuals in each bin using a scalar valued function. Then it plots the summaries $s_k$, for $k=1,\dots,b$, against $x_j$ (averaged within each bin). RIs can be obtained by simulating $l$ vectors of residuals from the model, binning and summarising them to obtain $\tilde{s}_k^v$, for $v=1,\dots,l$. Figure \ref{fig:diagGaus}e-f and \ref{fig:diagLss}f-g provide examples where the residuals are summarised using either the sample s.d. or the sample skewness. To extend this approach to 2D, \verb|l_gridCheck2D| uses the \verb|hexbin| package \citep{carr2011hexbin} to bin and summarise the observed and simulated residuals on a 2D grid of bins. We use hexagonal bins because of their favourable visual properties \citep{carr1987scatterplot}. The variability of the observed patterns can be taken into account by standardising the $s_k$'s using the mean and s.d. of the $l$ $\tilde{s}_k^v$'s in the same bin, as done to obtain Figure \ref{fig:diagLss}d-e. 

Binned residuals do not have to be reduced to scalar summaries. For example, Figure \ref{fig:panelPlots}a shows the output of \verb|l_gridCheck2D|, overlaid with a grid of worm-plots (detrended QQ-plots aiding visibility of deviations from the horizontal line \citep{buuren2001worm}), while Figure \ref{fig:panelPlots}b includes a sequence of k.d.e.s based on a coarser residual binning. The \verb|l_glyphs2D| layer summarises the residuals and the corresponding covariate values using any vector valued function, whose output can be rendered as a grid of glyphs. \cite{wickham2012glyph} point out that such glyph-maps are particularly useful for visualising spatio-temporal data, but here we show that they can be used also as residual checking tools. 

\subsection{Load forecasting: improving the Gaussian GAM} \label{sec:loadEx1}

\begin{figure} 
\centering
\includegraphics[scale=0.42]{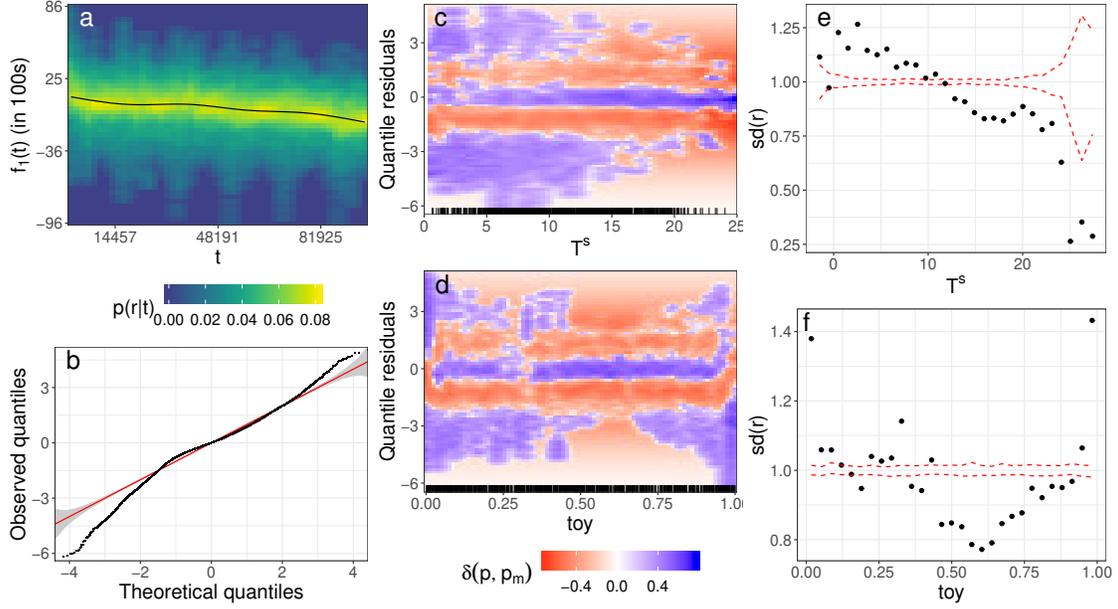} 
\caption{Gaussian GAM: a) $\hat{f}_1(t)$ and heatmap representing the conditional density of the partial residuals; b) normal QQ-plot with $99\%$ RIs; c) heatmap of $\delta_{p,p_m}=\{p(r|T^s)^{1/2} - p_m(r|T^s)^{1/2}\}^{1/3}$; d) same for $p(r|{\tt toy})$; e-f) dots are the s.d. of the binned observed quantile residuals and the dashed lines are $90\%$ RIs based on $l=50$ simulated residuals vectors.}
\label{fig:diagGaus}
\end{figure} 

We start assessing the adequacy of the Gaussian GAM by examining Figure \ref{fig:diagGaus}, which shows several diagnostics based on quantile residuals (we use this residual type throughout this application). The QQ-plot suggests that the residuals distribution $p(r)$ is fat-tailed and left-skewed, and the remaining plots provide more detail on model mis-specification. In particular, the partial residuals density heatmap in \ref{fig:diagGaus}a suggests the presence of a cyclical heteroscedastic component. This is confirmed by plot \ref{fig:diagGaus}d, which shows that demand is more variable in the winter than in the summer (${\tt toy} \approx 0.7$). The plot also shows that the demand distribution is left-skewed at year-end, which is not surprising, given that UK consumption drops in that period. Plot \ref{fig:diagGaus}c shows that demand is more variable at low temperatures, which is consistent with residential air conditioning being relatively uncommon in the UK. Plots \ref{fig:diagGaus}e and \ref{fig:diagGaus}f provide further evidence of heteroscedasticity along $T^s$ and ${\tt toy}$.

\begin{figure} 
\centering
\includegraphics[scale=0.4]{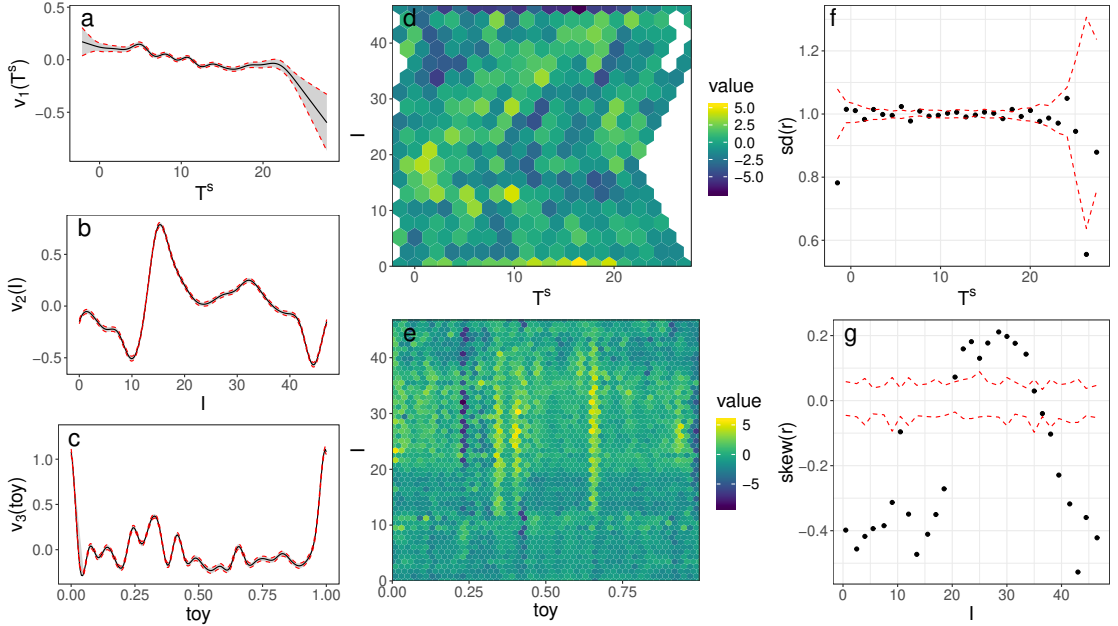} 
\caption{Location-scale GAM: a-c) smooth effects $\hat{v}_1(T^s)$, $\hat{v}_2(I)$ and $\hat{v}_3({\tt toy})$, with $95\%$ CIs; d) heatmap of s.d. of the observed quantile residuals in each bin, standardized using s.d. of $l=50$ residual vectors simulated from the model; e) same, but for sample skewness; f-g) analogous to Figure \ref{fig:diagGaus}e-f, but here plot g uses sample skewness.}
\label{fig:diagLss}
\end{figure}

Plots $1c$-$d$ provide much detail regarding the residual conditional distribution, and thus are useful for detecting residuals anomalies. However, an advantage of plots $1e$-$f$ is that they include useful information on the significance of the observed heteroscedastic pattern. Further, focusing on specific features of the residual distribution, such as the conditional variance, is helpful during the GAM model development process. Indeed, plots $1e$-$f$ provide a strong case in favour of including effects that allow the variance to vary smoothly with $T^s$ and ${\tt toy}$. Analogous plots (not shown) suggest that the variance varies also with some of the remaining covariates, hence we model the scale as follows
\begin{equation}\label{eq:varModel}
g\{\text{sd}(L_i)\} = \alpha_0h_{d(i)} + \sum_{j=1}^7 \alpha_j w^j_{d(i)} + v_1(T_i^s) + v_2(I_i) + v_3({\tt toy}_i),
\end{equation}
where $g(x) = \log (x - b)$ is a link function and $b>0$ is a small constant, included for computational stability reasons. The effects $v_2(I)$ and $v_3({\tt toy})$ are constructed using cyclic bases of rank 20 and 30, while $v_1(T^s)$ is based on a cubic spline basis of rank 20.

\begin{figure} 
\centering
\includegraphics[scale=0.33]{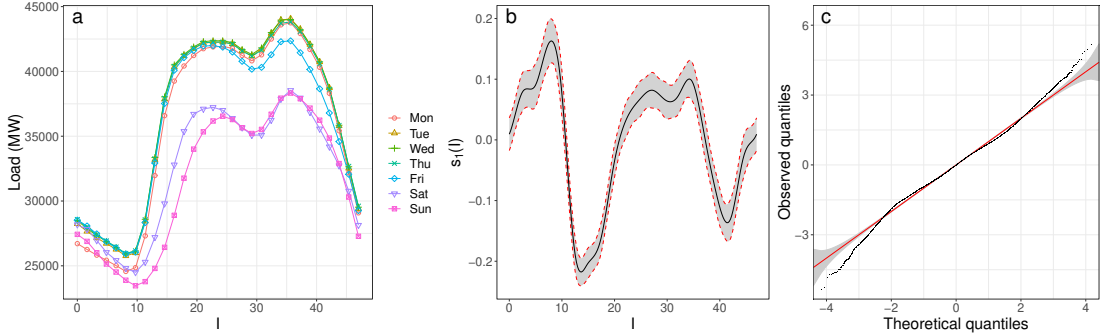} 
\caption{a) Daily load profiles, obtained by smoothing the load for each day; b) smooth effect $\hat{s}_1(I)$ with $95\%$ CIs; c) QQ-plot of quantile residuals for the shash GAMLSS model.}
\label{fig:profile}
\end{figure}

The AIC of this Gaussian location-scale model is $1.62\times 10^6$, while that of the basic GAM is $1.68\times 10^6$. Further, all terms in (\ref{eq:varModel}) have very low p-values ($< 10^{-6}$). The shape of $\hat{v}_1(T^s)$, shown in Figure \ref{fig:diagLss}a, implies that the variance decreases slightly with $T^s$, and \ref{fig:diagLss}f shows that the residual trend of \ref{fig:diagGaus}e has now disappeared. The effect of $I$ is stronger, with the conditional variance being maximal at peak times. It is likely that $\hat{v}_2(I)$ is adjusting for the fact that the shape of the daily load profile depends on the day of the week, as illustrated by Figure \ref{fig:profile}a, while our mean model (\ref{eq:gausModel}) includes a factor which simply shifts the profile depends on the day of the week. The discrepancy between the weekdays' and the Sunday's profiles reaches its peak around 8am, which is precisely the time at which $\hat{v}_2(I)$ is maximal. The fact that  $\hat{v}_3({\tt toy})$ increases sharply near year-end is partly due to the mean model not capturing the sudden decline in demand occurring during this period. The issue might be addressed by adopting an adaptive basis (see e.g. Section 5.3.5 of \cite{wood2017igam}), but this would lead to an overly complex model, as ${\tt toy}$ is part of a tensor product smooth.

The location model contains three bivariate smooths, hence it is reasonable to check for interactions acting on the conditional variance. Plot \ref{fig:diagLss}d shows one such check, which gives no clear evidence of a missing interaction in model (\ref{eq:varModel}). Figure \ref{fig:diagLss}e focuses on the residual skewness across ${\tt toy}$ and $I$. It shows a broad horizontal stripe of high skewness, between 10am to 7pm, and several thin vertical lines. The first pattern, as well as Figure \ref{fig:diagLss}g, suggests including a smooth effect modelling skewness along $I$. Instead, the vertical pattern is too irregular to be modelled via a smooth effect along ${\tt toy}$. Further, careful examination reveals that the vertical stripes correspond to variance peaks in plot \ref{fig:diagLss}c. While the variance and skewness patterns along ${\tt toy}$ could probably be reduced by more careful modelling of holidays in model (\ref{eq:gausModel}), further plots analogous to \ref{fig:diagLss}g (not shown) suggest the adoption of a GAMLSS model where the skewness depends on $I$, on the day of the week and on the holiday dummy variable. This will be described in Section \ref{sec:loadEx2}.

\section{Visualising smooth effect uncertainty} \label{sec:smoothEff}

Visualising the uncertainty of the fitted smooth effects is important for communicating the results of a GAM-based statistical analysis, but this is not trivial to do when dealing with multi-dimensional effects. Here we propose methods for visual assessment of the significance and uncertainty of bivariate smooths. The first approach consists in letting the heatmap's opacity be proportional to the significance of the fitted smooth at each location. Let $\hat{f}_{x_1 x_2} = \hat{f}(x_1, x_2)$ be the fitted smooth and $\hat{v}_{x_1 x_2} = \widehat{\text{var}}(\hat{f}_{x_1 x_2})$ be its estimated variance. We determine the opacity using
$
\alpha_{x_1 x_2} = t\{\Phi(|\hat{f}_{x_1 x_2}|/\sqrt{\hat{v}_{x_1 x_2}})\},
$
where $\Phi$ is a standard normal c.d.f. and $t(p):(0,1)\rightarrow(0,1)$ is a non-increasing transformation. Figure \ref{fig:sm1} uses $t(p) = \text{max}\{(1-z)^\gamma, \beta\}$, where $z = \text{max}(0, p-\delta)$, $\delta = 0.05$, $\gamma = 3$ and $\beta=0.2$. Varying the opacity allows identification of areas where the smooth effect is significantly different from zero, but it is not always effective for visualising the uncertainty of $\hat{f}_{x_1 x_2}$. This is better achieved by perturbing $\hat{f}_{x_1 x_2}$ using Gaussian white noise, with variance equal to $\hat{v}_{x_1 x_2}$. The result is that the heatmap's colours (see Figure \ref{fig:sm1}, or the Supplementary Material for a toy example) are proportional to the noisy function $\hat{g}_{x_1 x_2} = \hat{f}_{x_1 x_2} + z_{x_1 x_2}$, where $z_{x_1 x_2} \sim N\{0, \hat{v}_{x_1 x_2}\}$. One advantage of the methods just described is that significance and uncertainty are not presented in a binary `in-or-out' fashion, as is the case when using fixed significance levels or confidence bands, which helps conveying the meaning of statistical uncertainty. Further, the introduction of an extra dimension is avoided.

A different approach to visual uncertainty quantification is 3D rendering of the effects. \verb|mgcViz| offers this feature via the \verb|plotRGL| function, which uses the 3D interactive graphics offered by the OpenGL library \citep{neider1993opengl}, made accessible from \verb|R| by the \verb|rgl| package \citep{murdoch2001rgl}. Figure \ref{fig:smooth2} shows a snapshot of an \verb|rgl| graphic, which allows interactive manipulation (e.g. rotation) of each plot in the array. The plots use transparency to make both the fit and the confidence surfaces visible. 
Here interactivity is essential: such 3D objects are preferable to 2D equivalents only if they can be manipulated in real time.

\subsection{Load forecasting: adopting a GAMLSS model} \label{sec:loadEx2}

\begin{figure} 
\centering
\includegraphics[scale=0.4]{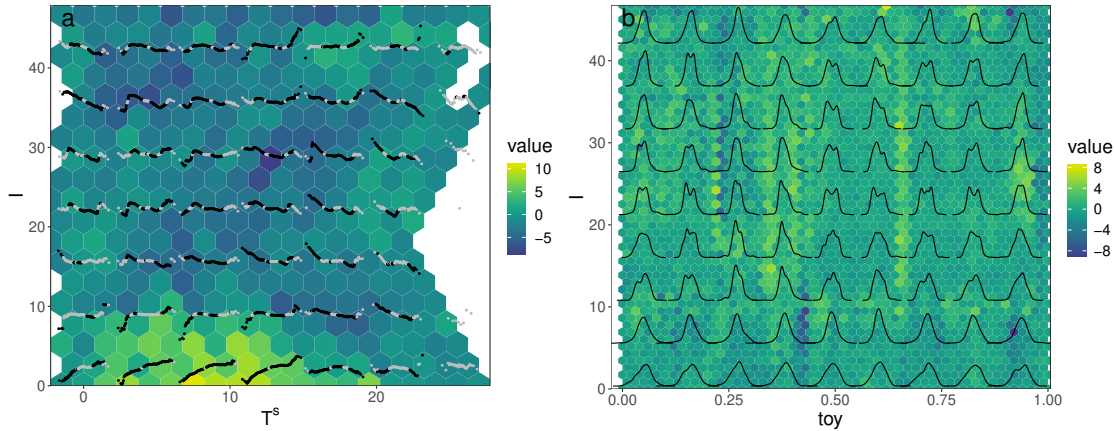} 
\caption{GAMLSS shash model: as in Figure \ref{fig:diagLss}d-e bin colours are computed using standardised s.d. (a) and skewness (b) in each bin. a) also shows several residual worm-plots, obtained using a coarser binning. Worms sections are black (grey) if they fall outside (inside) $95\%$ RIs. b) uses the binned residuals to compute a grid of k.d.e.s.}
\label{fig:panelPlots}
\end{figure}

To improve upon the location-scale model of Section \ref{sec:loadEx1}, we consider a GAMLSS model based on the sinh-arcsinh (shash) distribution of \cite{jones2009sinh}. We model its location, $\mu$, and scale, $\sigma > 0$, parameters using (\ref{eq:gausModel}) and (\ref{eq:varModel}), while for the skewness we use
\begin{equation}\label{eq:skewModel}
\epsilon_i = \gamma_0h_{d(i)} + \sum_{j=1}^7 \gamma_j w^j_{d(i)} + s_1(I_i),
\end{equation} 
%
where $s_1(I_i)$ is a smooth effect, constructed using a cyclic spline basis of rank  20. The shash model contains also a parameter, $\delta>0$, controlling the tail behaviour, which we were unable to identify. In particular, $\delta$ was diverging toward high values, where the density becomes insensitive to its value \citep{jones2009sinh}. Hence, we preferred setting $\delta=1$ (Gaussian-like tails), which leads to a model containing 1013 regression coefficients and 11 smoothing parameters. The increase in complexity seems justified, as the AIC of the shash model is $1.608\times 10^6$ and all of the terms in (\ref{eq:skewModel}) are significant at 0.01 level.

Figure \ref{fig:profile}b shows that the shape of $\hat{s}_1(I)$ is roughly consistent with the skewness pattern observed in Figure \ref{fig:diagLss}e. The QQ-plot in Figure \ref{fig:profile} is much improved relative to the one shown in Figure \ref{eq:gausModel}, especially in the lower tail. Fitting the four-parameters shash density to the residuals of the shash GAMLSS model returns $\hat{\delta}\approx 1$ and an almost identical QQ-plot, suggesting that we have not lost much by fixing $\delta$. Still, the QQ-plot indicates that the fit could be improved further. Figure \ref{fig:panelPlots} provides more evidence of this. In particular, the worm-plots in \ref{fig:panelPlots}a show large deviations of quantile residuals from normality, particularly in the lower tail. Further, the heatmap shows that the residuals are over-dispersed between midnight and 2am but not before midnight, which suggests that using a cyclic basis for $v_2(I)$ might not be appropriate. The binned k.d.e.s in Figure \ref{fig:panelPlots}b give evidence of multimodality, which might be attributable to the shape of daily load profile being different depending on day of the week and to the fact that our model does not integrate special tariff information.

\begin{figure} 
\centering
\includegraphics[scale=0.4]{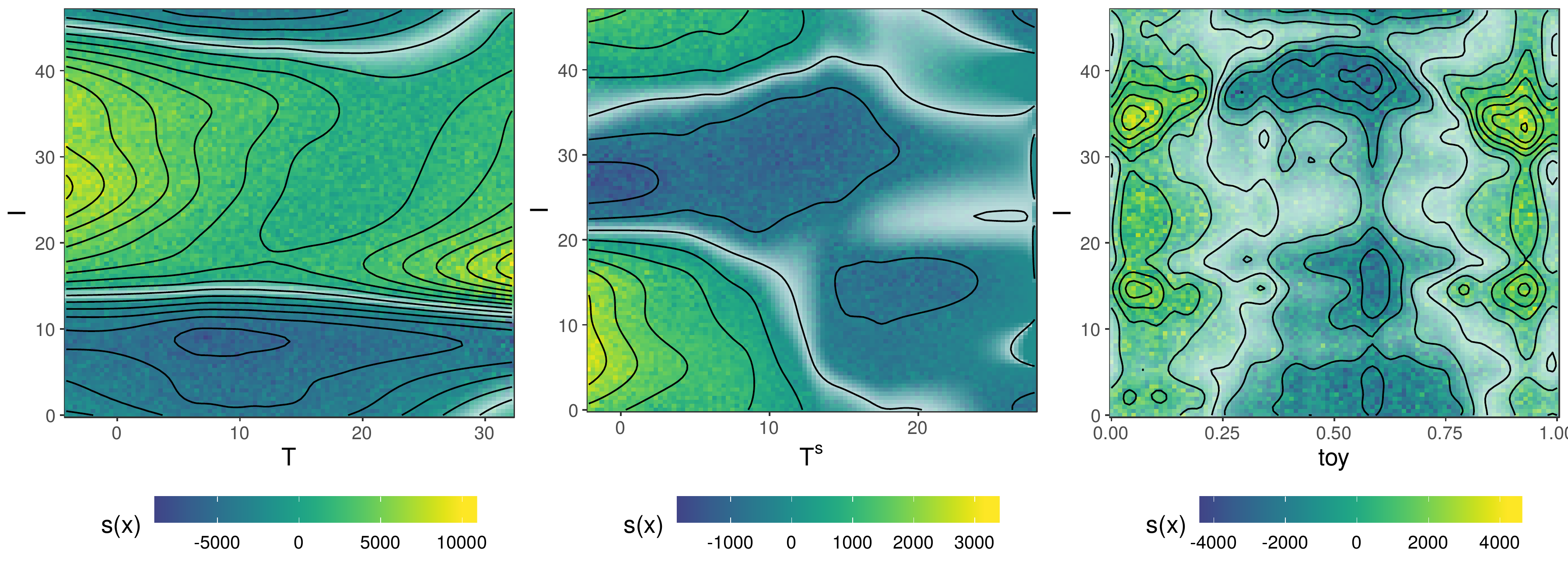} 
\caption{GAMLSS shash model: smooth effects $\hat{f}_1(T, I)$, $\hat{f}_2(T^s, I)$ and $\hat{f}_3({\tt toy}, I)$. Randomisation and transparency quantify their uncertainty and significance.}
\label{fig:sm1}
\end{figure}

\begin{figure}
\centering
\includegraphics[scale=0.4]{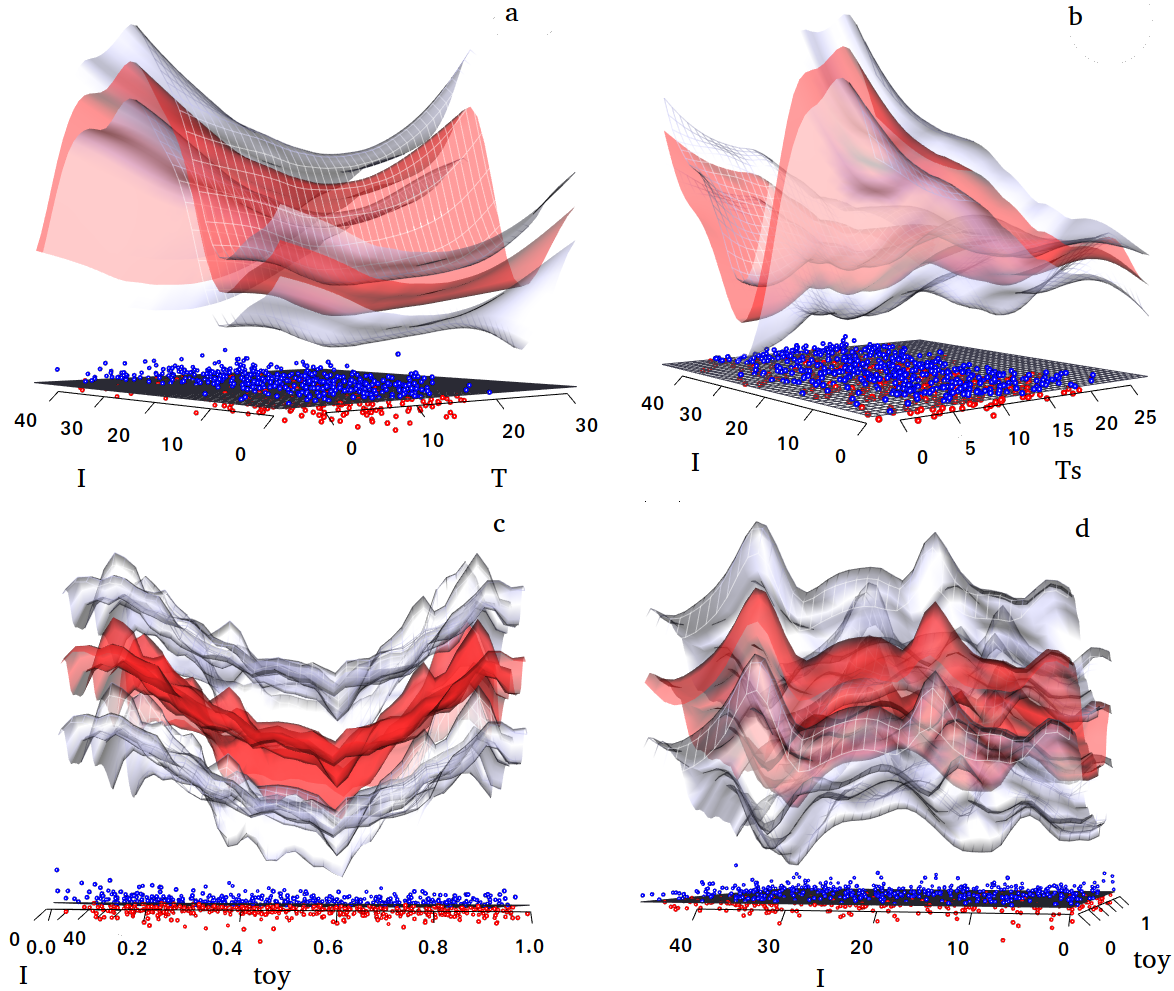} 
\cprotect\caption{GAMLSS shash model: snapshot of an \verb|rgl| graphic showing a) $\hat{f}_1(T, I)$, b) $\hat{f}_2(T^s, I)$ and c-d) $\hat{f}_3({\tt toy}, I)$. The fitted effects are sandwiched between the $66\%$ confidence surfaces, which are showed in light grey. Each plot includes a sub-sample of the residuals, obtained using random sampling with replacement. $\hat{f}_3({\tt toy}, I)$ is presented from two viewpoints, which show that this effect is smoother across $I$ than ${\tt toy}$.}
\label{fig:smooth2}
\end{figure}

Figure \ref{fig:sm1} shows the bivariate smooth effects for the location parameter $\mu$. As expected the effect of variations in the instantaneous temperature $T$ is much stronger during the day, due to manual heating regulation. In contrast, low $T^s$ has a strong positive effect at night, probably because of storage heaters. Notice that the effect $\hat{f}_2(T^s, I)$ is barely significant for $T^s>20$, as UK temperatures rarely stay much above $20^\circ$C for several consecutive days. The effect $\hat{f}_3({\tt toy}, I)$ is quite complex and it is characterised by higher uncertainty. It shows four maxima, corresponding to daily peak times, separated along ${\tt toy}$ by the year-end demand drop. Figure \ref{fig:smooth2} shows the same effects in three dimensions. 

\section{Conclusions}

We presented a set of scalable visual tools meant to facilitate results presentation, model checking and building for general GAMs. In the example we emphasised the use of visual aids for interactive model building, because we feel that this approach is much preferable to automated variable selection approaches when dealing with large data sets and complex models. Further, such visual checks allow practitioners to understand why an effect was included and thus to develop more confidence in the chosen model. This is key to fostering the adoption of more sophisticated GAM models in large industrial institutions, such as \'Electricit\'e de France, where forecasting errors have major practical consequences.

The object-oriented layer-based framework implemented by \verb|mgcViz| aims at facilitating future extensions of the visualisation methods proposed here. In particular, while the package already contains diagnostic layers that are specific to quantile GAMs \citep{fasiolo2017fast}, we plan to develop bespoke methods for other non-standard models, such as functional GAMs \citep{mclean2014functional}. More demanding extensions would be providing general methods for creating animated version of current plot types, which would be useful for smooth effect uncertainty visualisation \citep{bowmanvisual}, and tools for comparing plots generated under different GAM models. The latter development would be useful for model comparison purposes, particularly in conjunction with new plots focusing on the predictive performance, rather than the goodness-of-fit, of the models involved in the comparison.

%
%

\section*{Acknowledgements}

We thank three anonymous reviewers for a large number of useful comments on an earlier draft, and Fabian Scheipl for suggestions which helped us improving \verb|mgcViz|. This work was funded by EPSRC grants EP/K005251/1, EP/N509619/1 and by EDF. 

\bibliography{biblio.bib}

\newpage

\begin{center}
{\large\bf Supplementary material for ``Scalable visualisation methods for modern Generalized Additive Models''}
\end{center}

\renewcommand{\appendixpagename}{}
\begin{appendices}

\renewcommand{\theequation}{S\arabic{equation}}

\setcounter{equation}{0}

\cprotect\section{Interpreting the output of \verb|l_densCheck|}

This section contains few examples of diagnostic plots produced by the \verb|l_densCheck| layer, meant to introduce practitioners to this new type of visualization. Throughout the section we will be using quantile residuals, $\Phi^{-1}\{F_m(y|{\bf x})\}$, hence the reference theoretical density (i.e. the expected density under a well-specified model) is a standard normal density. In each example, we simulate $10^4$ responses using the sinh-arcsinh of \cite{jones2009sinh} and we fit a sequence of Generalized Additive Models for Location Scale and Shape (GAMLSS) \citep{rigby2005generalized} based on this density, where the linear predictor controlling one of the sinh-arcsinh model parameters is misspecified.

\begin{figure}
\centering
\includegraphics[scale=0.55]{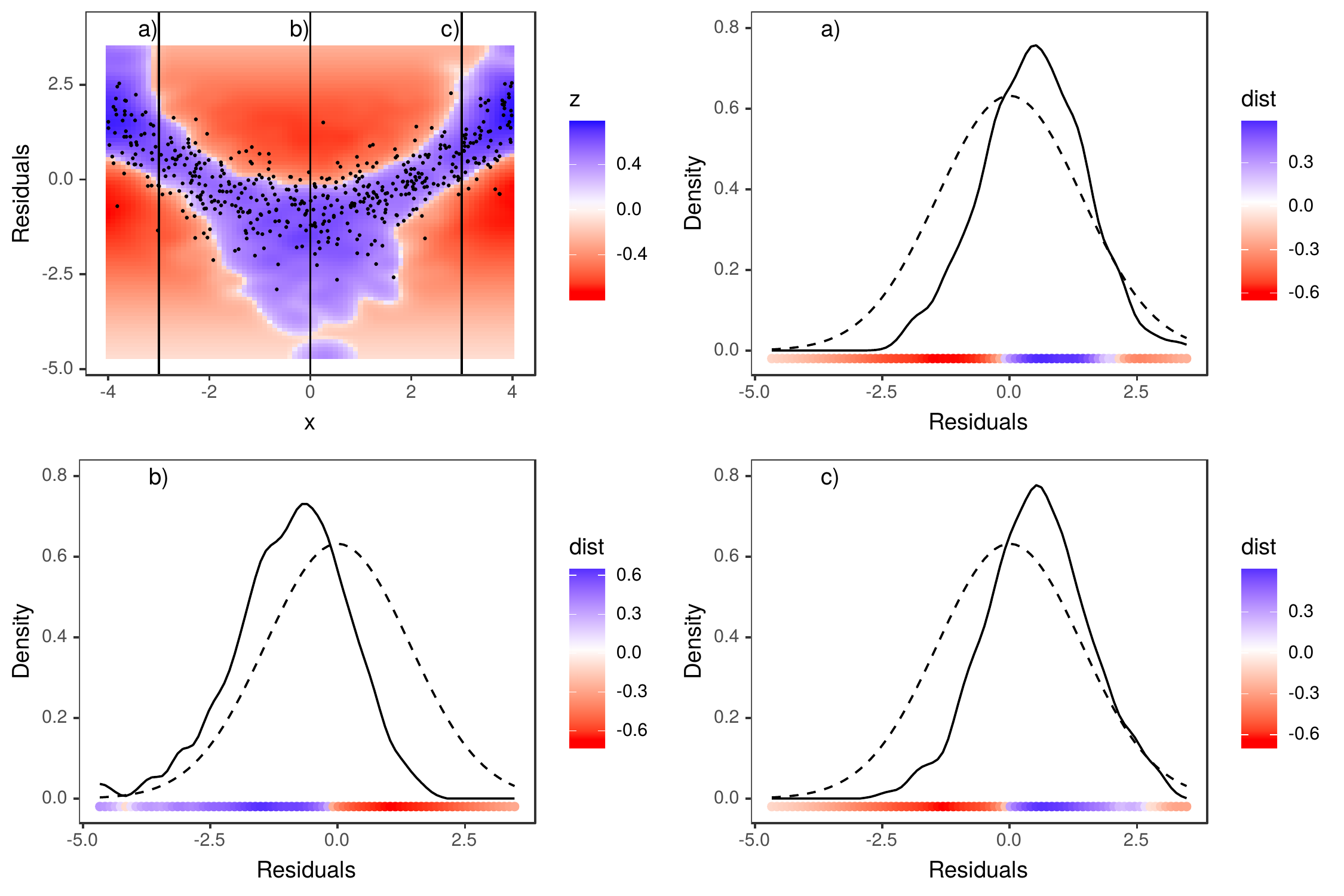} 
\cprotect\caption{Misspecified mean model: output of \verb|l_densCheck| (top-left) and three density plots comparing the theoretical residuals density, $N(0,1)$ (dashed), with three estimates of the conditional residuals density, $p(r|x)$ (solid), with $x$ fixed at locations a), b) and c). The colours of the points at the bottom of a), b) and c) match those in the corresponding slice of the heatplot on the top-left. The black points on the top-left plot are a sub-sample of 500 residuals.}
\label{fig:densCheckMean}
\end{figure}

The plot on the top-left of Figure \ref{fig:densCheckMean} shows the output of \verb|l_densCheck| in a case where the location (or mean) model is misspecified. In particular, the residuals (a sub-sample of which is represented by the black points) show a systematic quadratic pattern in their mean, when ordered using the values of the covariate $x$. The remaining three plots compare kernel estimates of the conditional density of the residuals, $p(r|x)$, with the reference $N(0,1)$ density, for three values of $x$. Notice that the heatmap is blue (red) when the empirical density is lower (higher) than the theoretical density. This plot suggests that the linear predictor for the location should include a smooth effect of $x$.

\begin{figure} 
\centering
\includegraphics[scale=0.55]{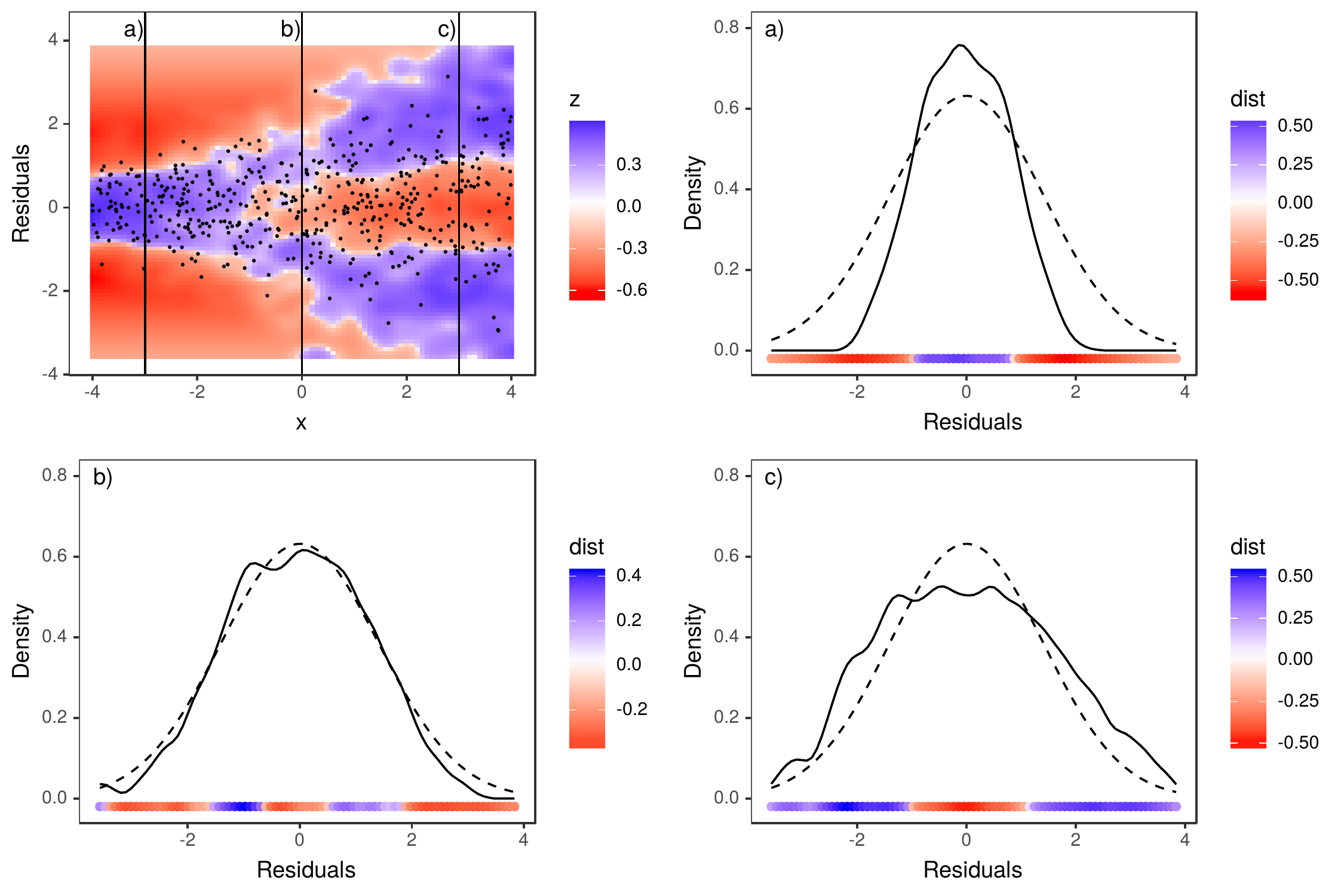} 
\cprotect\caption{Misspecified variance model: all plotted elements have the same interpretation as in Figure \ref{fig:densCheckMean}, but here the misspecification is in the conditional variance, not the mean.}
\label{fig:densCheckVar}
\end{figure}

Figure \ref{fig:densCheckVar} shows another example where the conditional variance, not the mean, of the residuals varies with $x$. In particular the variance increases with $x$, in fact the estimated $p(r|x)$ is under-dispersed for $x\approx -3$, has roughly the correct variance for $x\approx 0$ and is over-dispersed for $x\approx 3$. In the context of GAMLSS modelling this plot suggest that the linear predictor for the scale should include a smooth effect of $x$. 

\begin{figure} 
\centering
\includegraphics[scale=0.55]{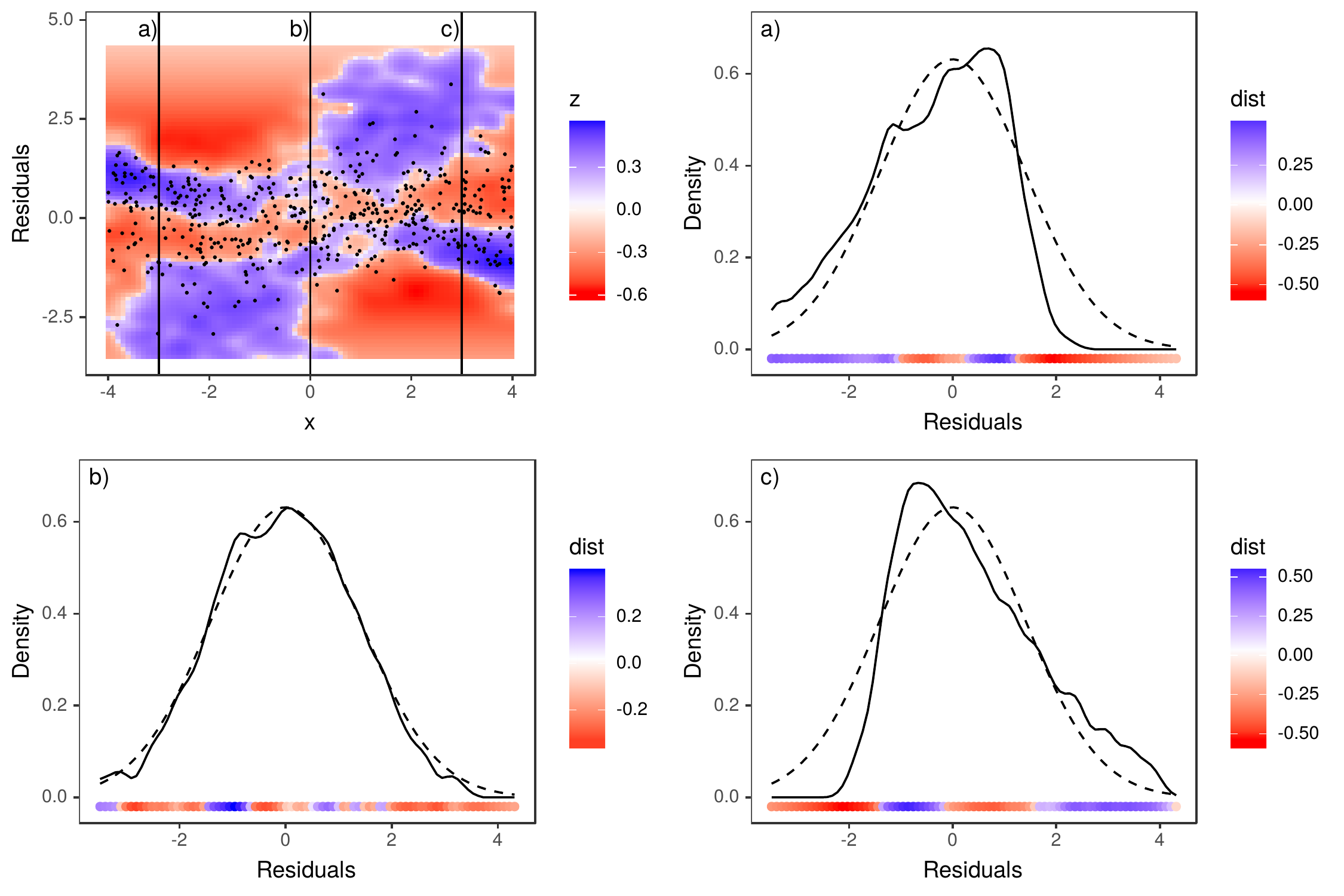} 
\cprotect\caption{Misspecified skewness model: all plotted elements have the same interpretation as in Figure \ref{fig:densCheckMean}, but here the misspecification is in the conditional skewness, not the mean.}
\label{fig:densCheckSkew}
\end{figure}

\begin{figure} 
\centering
\includegraphics[scale=0.55]{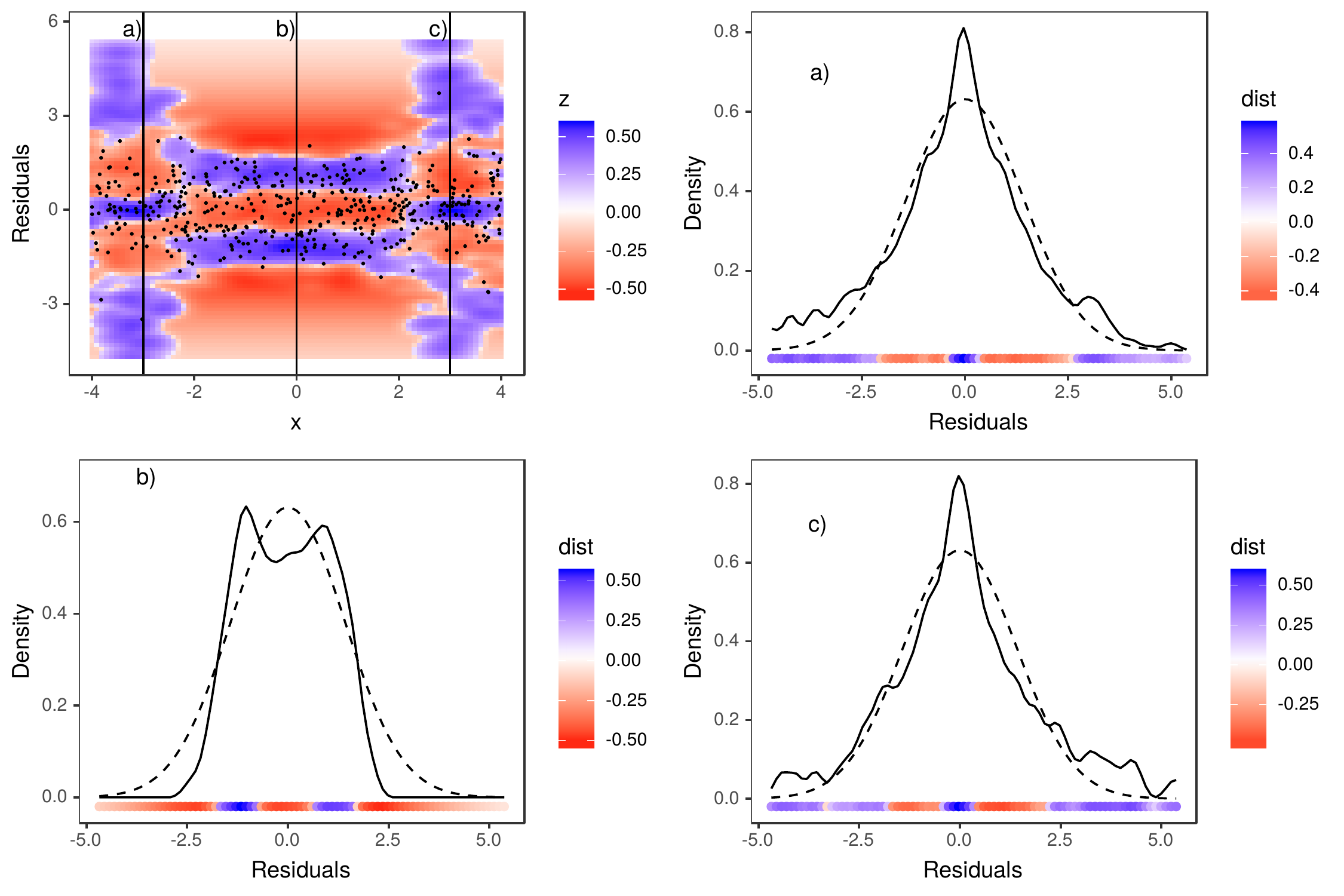} 
\cprotect\caption{Misspecified tail model: all plotted elements have the same interpretation as in Figure \ref{fig:densCheckMean}, but here the misspecification is in the conditional kurtosis, not the mean.}
\label{fig:densCheckKurt}
\end{figure}

Figure \ref{fig:densCheckSkew} shows a similar sequence of plots where the skewness of the residuals varies with $x$. In particular, the residuals density is skewed to the left (right) at $x\approx -3$ ($x\approx 3$), while it is approximately symmetric for $x\approx 0$. Here a smooth effect for $x$ is missing from the linear predictor controlling the skewness of the response density. Figure \ref{fig:densCheckKurt} gives an example where the kurtosis or weight of the tails is higher than expected under the model for $x \approx \pm3$ and too low for $x \approx \pm 0$. Notice that the heatmap shows three blue modes for $x \approx \pm3$, while the heatmap in Figure \ref{fig:densCheckVar} shows only two blue modes at $x \approx \pm 3$. This fact allows us to distinguish excessively heavy tails from over-dispersion. 

\begin{figure} 
\centering
\includegraphics[scale=0.40]{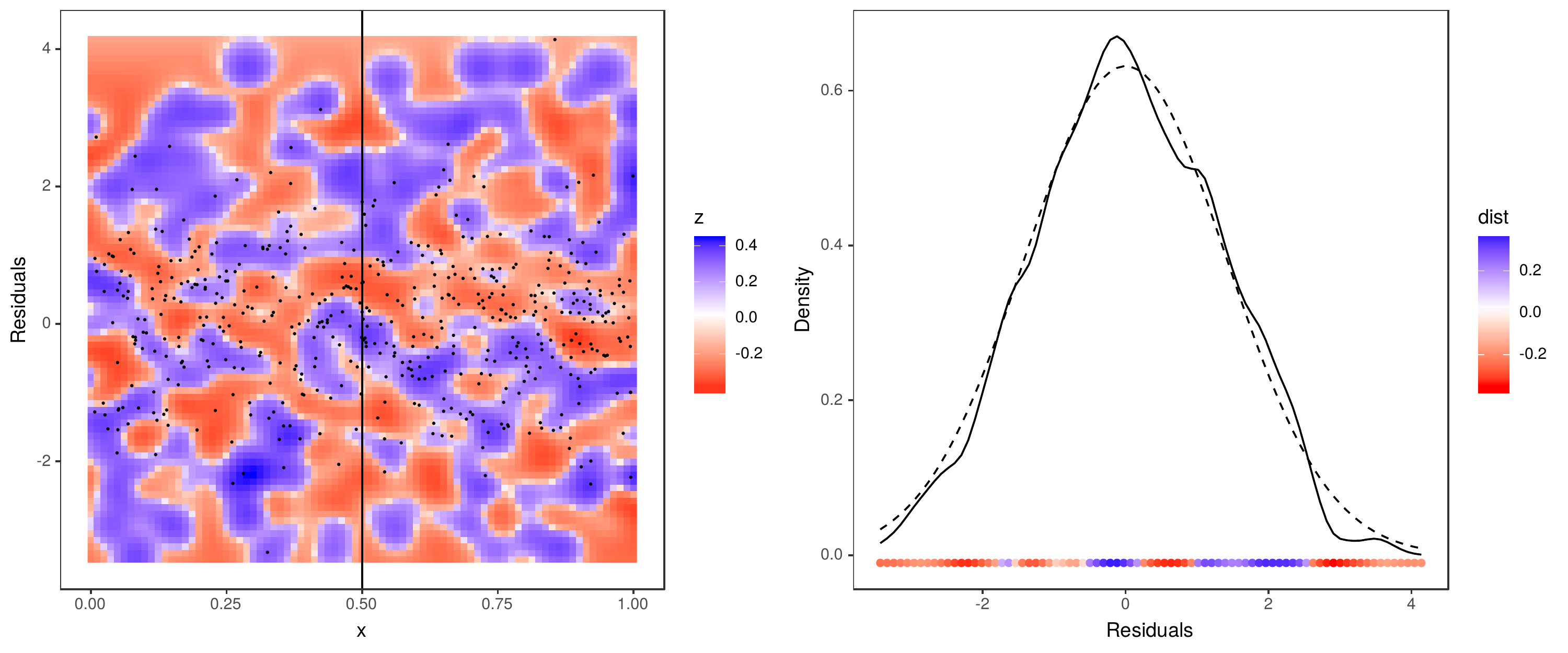} 
\cprotect\caption{Well specified model: all plotted elements have the same interpretation as in Figure \ref{fig:densCheckMean}, but here the model is well specified and no clear pattern is visible.}
\label{fig:densCheckWell}
\end{figure}

Figure \ref{fig:densCheckWell} shows the output of the \verb|l_densCheck| layer for an example where the response distribution is well-specified. Here no clear residual patterns is visible, and the empirical density differs from the theoretical $N(0, 1)$ density in a random manner. 

\section{Uncertainty visualisation in 2D smooth effect plots}

\begin{figure} 
\centering
\includegraphics[scale=0.6]{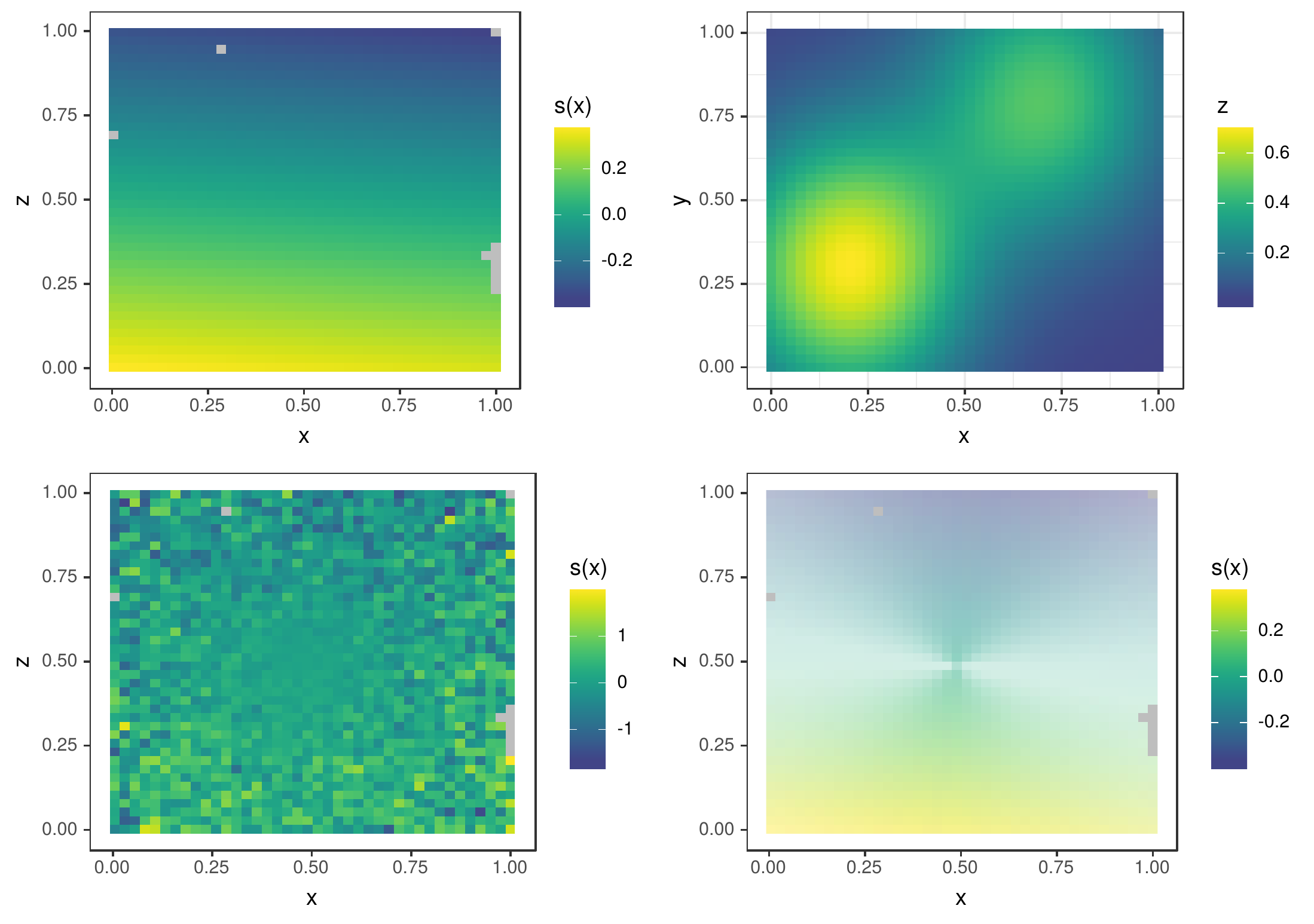} 
\caption{Fitted 2D effect $\hat{f}_{xy}$ (top-left), truth $f_{xy}$ (top-right), perturbed fitted effect $\hat{g}_{x z}$ (bottom-left) and fitted effect plot where the opacity is proportional to the significance of $\hat{f}_{xy}$ (bottom-right).}
\label{fig:smoothUncert}
\end{figure}

\begin{figure} 
\centering
\includegraphics[scale=0.6]{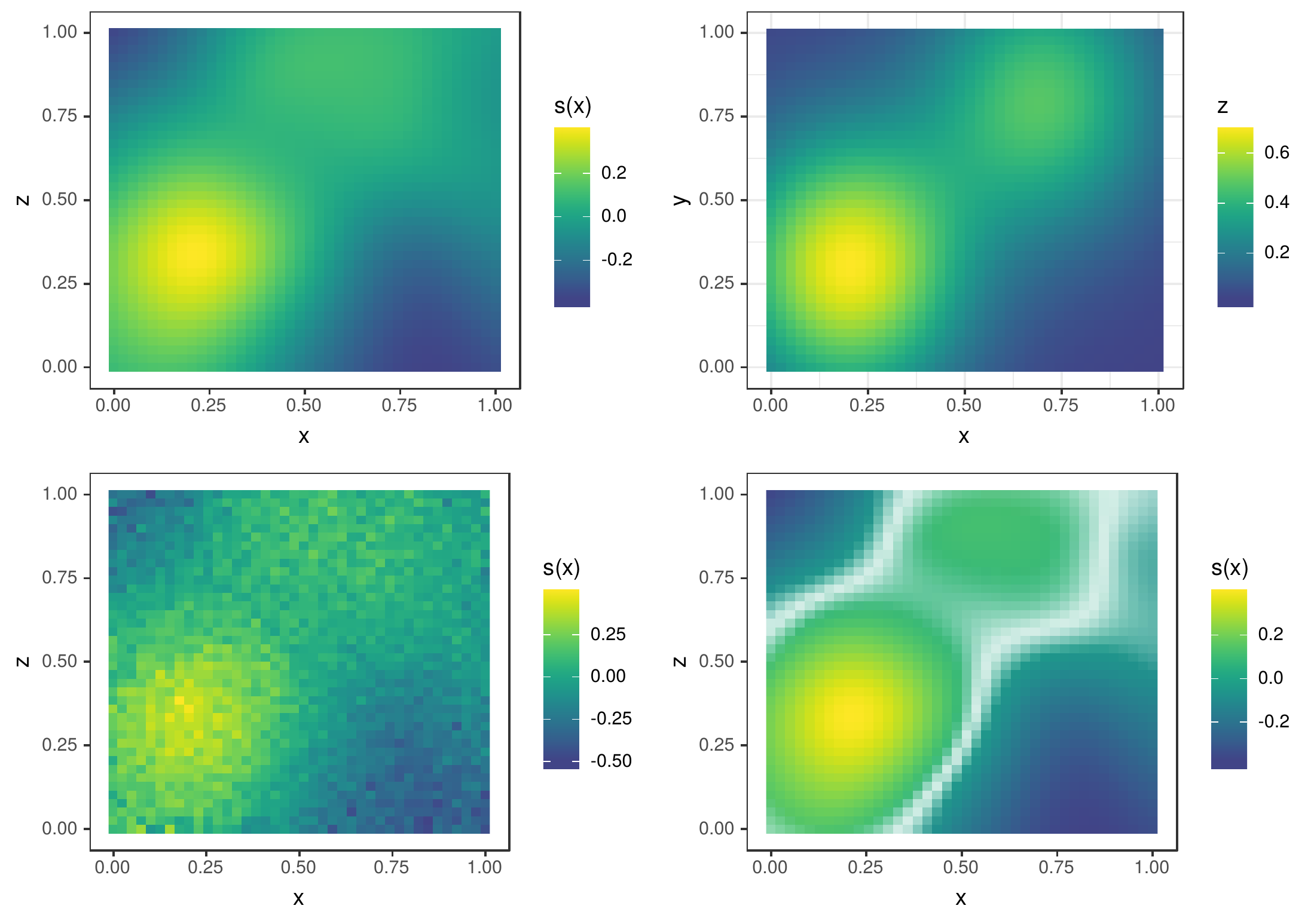} 
\caption{Same as Figure \ref{fig:smoothUncert}, but with $n = 10^5$.}
\label{fig:smoothUncert2}
\end{figure}

Here we use a simple example to demonstrate some of the  methods for representing the uncertainty of fitted two dimensional smooth effects described in the main paper. In particular, we simulate $n = 200$ data points from 
$$
y=f(x,z)+\epsilon, \;\;\; \epsilon \sim N(0,\sigma^2=4),
$$ 
where $x$ and $z$ are $\text{U}(0, 1)$ distributed. We fit the data using a Gaussian GAM model with a tensor product smooth for $f_{xy} = f(x,z)$, formed by 25 basis function. Figure \ref{fig:smoothUncert} shows $f_{xy}$, its tensor product estimate, $\hat{f}_{xy}$, a perturbed version of the latter, $\hat{g}_{x, z}$, and a heatmap of $\hat{f}_{xy}$ where the opacity is proportional to the significance of $\hat{f}_{xy}$ (as defined in the main text). The first plot of $\hat{f}_{xy}$ shows that the smoothing penalty has shrunk the tensor product effect to a flat surface, with a linear gradient wrt $z$. However, the perturbed heatmap does not show any linear effect but only white noise, and the heatmap is almost transparent when varying opacity is used. Hence both smooth effect uncertainty visualisation methods suggest that the vertical gradient in $\hat{f}_{xy}$ is purely random. Figure \ref{fig:smoothUncert2} shows the same plots for $n=10^5$. When such a large sample size is used (the signal-to-noise ratio is quite low in this example), the same smooth pattern is clearly visible in all plots.   

\end{appendices}

\end{document}